# Low temperature crystallization of diamond-like carbon films to graphene by low energy plasma surface treatment

S.S. Tinchev[a]

Institute of Electronics, Bulgarian Academy of Sciences, Sofia 1784, Bulgaria

Abstract:

Plasma surface modification was used to fabricate graphene on the top of insulating diamond-like carbon films. It is shown that by a combination of pulsed argon plasma treatment and thermal annealing at 350°C is possible to achieve partial crystallization of amorphous carbon to graphene. The observed Raman spectra are typical for defected graphene – splitted D- and G-peaks and a broad 2D-peak. This result is very encouraging and we hope that by improving this technology it will be possible to fabricate defect-free graphene, which can be used in electronics without transfer to other substrates.

[a] Electronic mail: stinchev@ie.bas.bg

Graphene, the first 2D atomic crystal ever know possess interesting electrical properties, especially high carrier mobility, which promise its bright future in electronics. However, there is still no suitable technology for fabrication of graphene for general electronic applications. Today the most successfully fabrication technology is CVD on metals followed by a transfer process on insulated substrates as needed usually in electronics. This process is quite successful – up to 30-inch graphene sheets were fabricated and successfully transferred to plastic substrates [1]. Although such technology is suitable for transparent electrodes, a general applicable technology for fabrication of graphene on insulating substrates is needed. Here one should mentioned that the fabrication of graphene from SiC by thermal decomposition [2] is not an alternative because the high temperatures ( $\geq$ 1300°C) needed to sublimate Si atoms make this technology incompatible with the existing silicon electronics.

Recently [3] we proposed an idea for fabrication of graphene on the top of insulating amorphous carbon films by low-energy ion modification. In this low-temperature process the surface of the amorphous carbon could crystallize to graphene as a result of point defect creation and enhanced diffusion caused by the ion bombardment. Different ions can be used to modify diamond-like carbon films, for example carbon and hydrogen ions as inherent to the starting material (a-C:H) in our experiments. We choose argon ions, which are widely used in the microelectronic technology and as a noble gas should not react with the carbon. To estimate the necessary energy and doses of the ions in order to modify only some monolayers on the surface of the amorphous carbon films the Monte Carlo SRIM-2008 program [4] was used. We found [3] that for energy of the argon ions 1 keV the dose should be ~ $4.5 \times 10^{15}$ Ar$^+$/cm$^2$ in order to break all bonds of the surface of the amorphous carbon films. This estimated value is in good agreement with the experimental value found in the literature [5].

In our first experiments [3] the films were modified in DC magnetron system at unipolar pulsed discharges. Pulse biasing of the magnetron is needed because the diamond-like carbon films are highly insulating and ion bombardment with DC voltage would cause charging of the film. The reason for use magnetron in the film modification was the possibility for easy production of argon ions with low energy and high ion density. However, this choice has also a drawback. To achieve the necessary dose of about $4.5 \times 10^{15}$ Ar$^+$/cm$^2$ it was found that the modification time should be shorter then 1 s. Although these first experiments were quite successful, the short modification time was difficult to control and to vary in order to optimize the technology.

Therefore another system was built later. It is a simple diode system. Both electrodes are cooper strips fixed on an alumina substrate and the samples could be placed on/or near the cathode. The results presented in this paper were obtained for samples placed directly on the cathode. The cathode and samples were not cooled because to our estimations it should not be significant heating during the ion bombardment. The voltage amplitude was 400 V, pulse frequency of 66 kHz and pulse time of 10 μs. The system was evacuated by a combination of diffusion and a mechanical pump. During the modification the pressure of the chamber was $3 \times 10^{-1}$ Torr.

Fig. 1 shows the calculated profile of the vacancies produced in amorphous carbon films for 400 eV argon ion modification. As expected only 1 nm of the surface of the film (about 3 monolayers) will be modified. One can see in Fig. 1 also the profile of the implanted argon ions – the curve "Ion range" with maximum at about 1.6 nm. Obviously the argon ions are implanted far behind the modified surface region and they should not introduce additional effects. The calculated necessary dose for this ion energy is nearly the same as calculated in [3], about $4.5 \times 10^{15}$ $Ar^+/cm^2$.

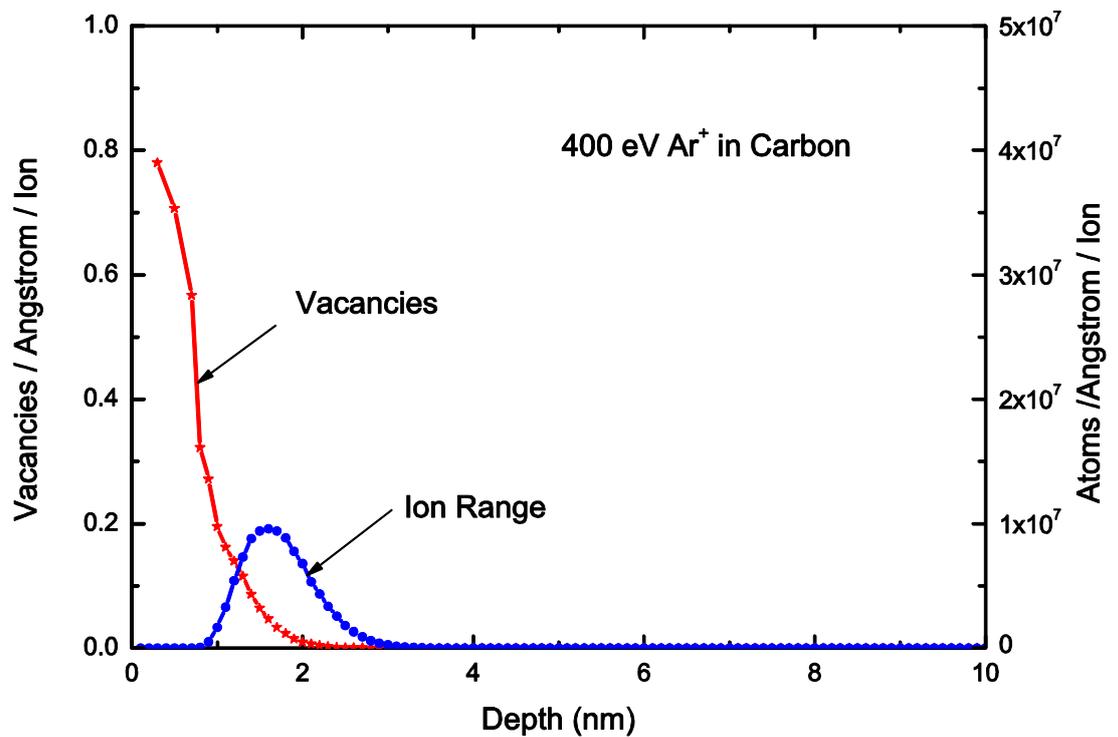

Fig. 1. Calculated vacancies profile and argon ion distribution for 400 eV $Ar^+$ plasma treatment of amorphous carbon.

Here one significant remark should be made. Usually in plasma-based implantation systems (as actually our system is) the ion dose incident in the sample is measured by integrating the current induced in the sample by the ion beam. However, because our samples are highly insulating, there is problem to measure the sample current. Due to the difficulties associated with dose measurements, in this paper we will give only the irradiation times instead of the ion doses.

Raman spectra of the samples were measured before and after sample modification at room temperature. Raman spectra were obtained using laser with 633 nm wavelength and laser power smaller then 0.9 mW in order to prevent modification of the films by the laser irradiation. Raman spectra of the plasma treated (but not annealed) samples show only small differences in comparison with the non implanted samples. This can be expected because at room temperature the diffusion coefficient of the carbon is not high enough for its crystallization. Obviously an annealing should be made after the plasma treatment. Fig. 2 shows Raman spectra of the samples after annealing at 350°C for 6 hours. This temperature was chosen as maximal temperature, which cannot affect the underplaying amorphous carbon. In Fig. 3 it is demonstrated that without plasma treatment an annealing of a sample does not change significant the Raman spectrum. In our first experiments [3] the annealing temperature was some lower - 300°C. Later we have found that it can be increased to 350°C.

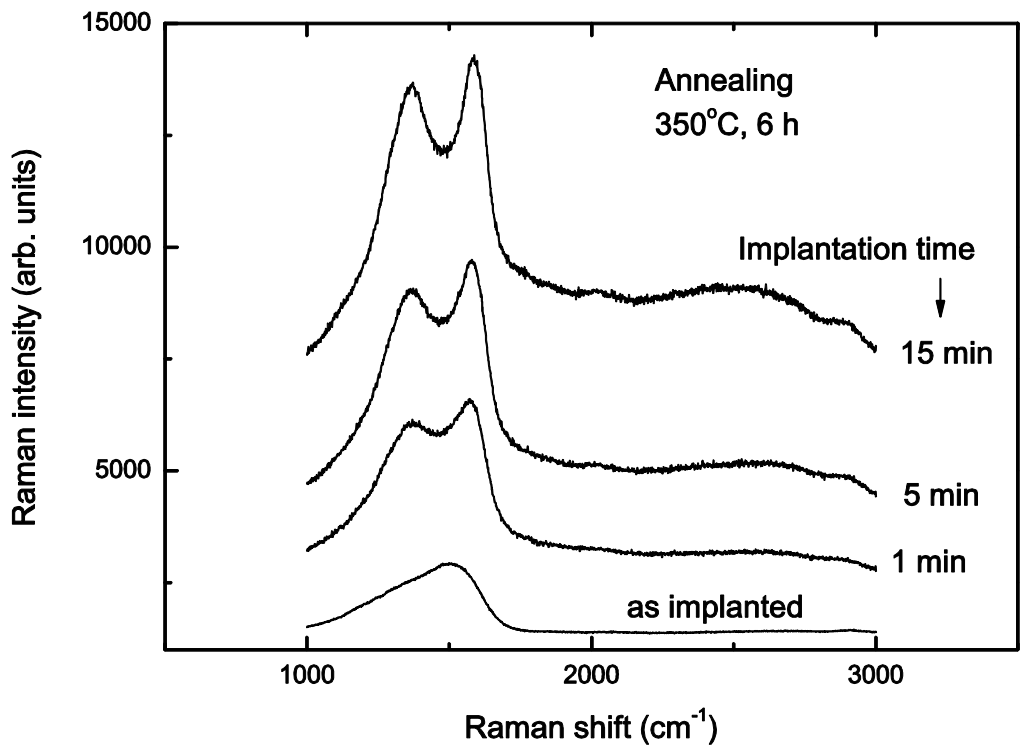

Fig. 2. Raman spectra of samples modified for different times and annealed at 350°C for 6 hours.

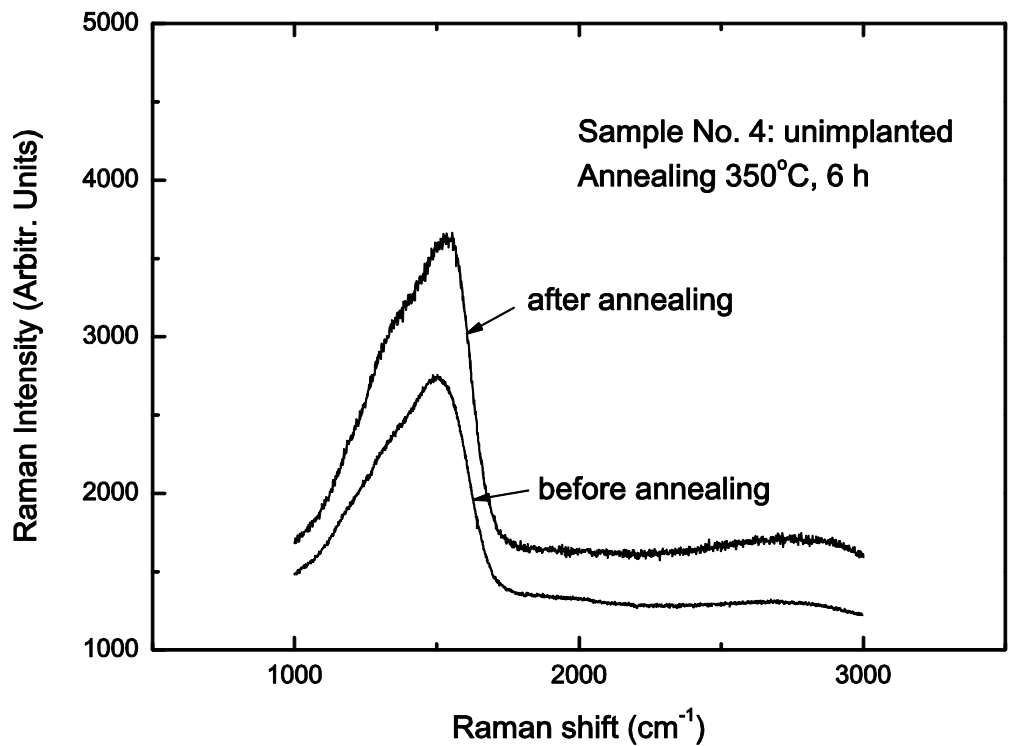

Fig. 3. Raman specra of unmodified sample before and after annealing at 350°C for 6 hours.

Obviously the most significant changes of the Raman spectra after annealing were the splitting of the broad amorphous spectrum into two distinct D- and G-peaks. This is typical for partial crystalline carbon with small crystalline size. Without plasma treatment such partial crystallization of amorphous carbon can be achieved by annealing at temperatures 800–900°C. It is well known that in the graphite or graphene presence of the D peak indicates the presence of disorder. However, in amorphous carbons the development of a D peak indicates ordering. The D-peak intensity rises with increasing plasma treatment time, i.e. the crystallization is enhanced. However, its position remained stable, as Fig. 4 presents. In contrast to this, the G-peak shows an upshift ( ~ 12 $cm^{-1}$) from 1572 $cm^{-1}$ to 1584 $cm^{-1}$, close to the typical G-peak of graphene. As the plasma treatment time increased, the G-peak narrowed, indicating again enhanced ordering (crystallization) of the modified amorphous carbon.

The $I_D/I_G$ ratio widely used for characterization of carbon materials is shown also there. In this paper we refer to $I_D/I_G$ as the ratio of peak heights. With increasing implantation time this ratio rises up to ~ 0.96 indicating $sp^2$ cluster size of about 1.5 nm [6]. However, such conclusions should be made very carefully because interpretation of Ramen signal of such complicated system like graphen on amorphous carbon is not easy.

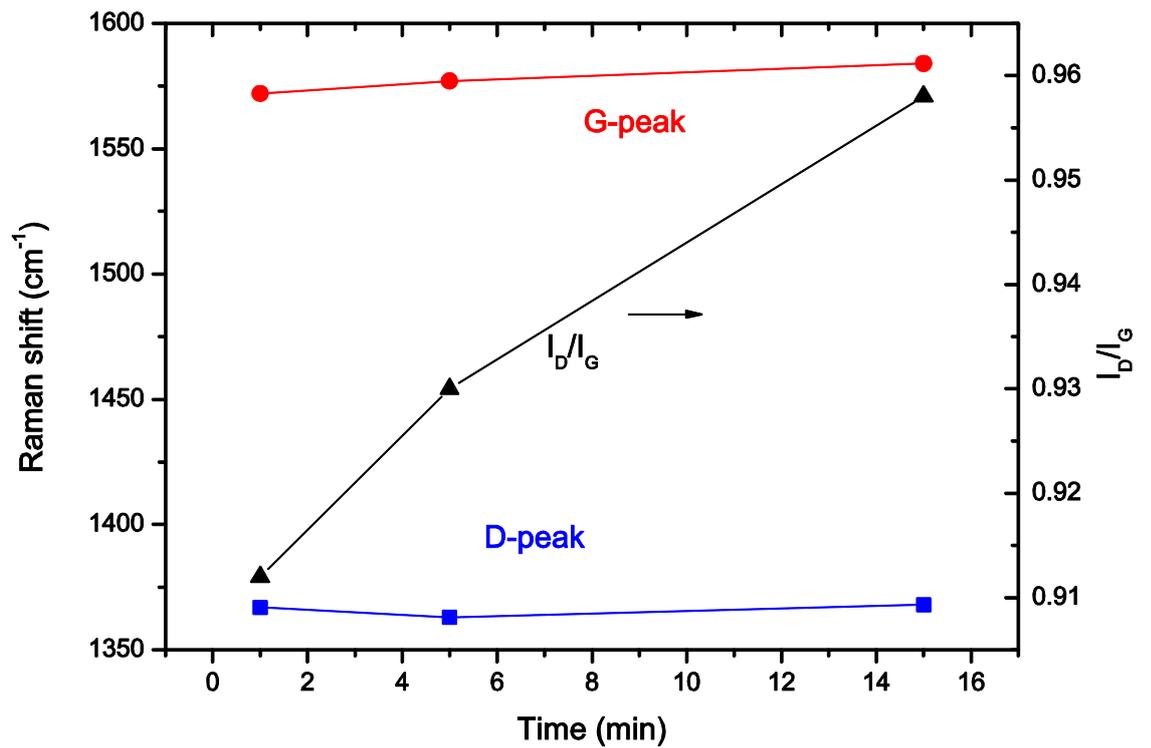

Fig. 4. D-peak and G-peak positions vs. plasma treatment time. Sown is here also the intensities ratio of D-peak and G-peaks.

The single and sharp second order Raman band (2D) is widely used as a simple and efficient way to confirm the presence of single layer graphene. In the Raman spectra of our samples in the 2D band region one can recognize a peak at 2700 cm$^{-1}$ which is however, not sharp as expected. These broad Raman spectra are typical for defected graphene [7-9]. With increasing modification time (ion dose) this peak rises and two other small peaks at about 2000 cm$^{-1}$ and 2900 cm$^{-1}$ can be seen. The peak at 2900 cm$^{-1}$ can be identified as D + D' combination [7]. At the moment the origin of the peak at about 2000 cm$^{-1}$ is not clear.

In conclusion surface modification by low energy pulsed argon plasma was used to fabricate graphene on the top of insulating diamond-like carbon films. It is shown that by following low temperature thermal annealing at 350°C is possible to achieve partial crystallization of amorphous carbon to graphene. The observed Raman spectra are typical for defected graphene – splitted D- and G-peaks and a broad 2D-peak. This result is very encouraging and we hope that by improving this technology it will be possible to fabricate defect-free graphene, which can be used in electronics without transfer to other substrates.

Acknowledgement:

The authors would like to thank Dr. Evgenia Valcheva for the Raman measurements.

References:

1. Sukang Bae, Hyeongkeun Kim, Youngbin Lee, Xiangfan Xu, Jae-Sung Park, Yi Zheng, Jayakumar Balakrishnan, Tian Lei, Hye Ri Kim, Young Il Song, Young-Jin Kim, Kwang S. Kim, Barbaros Ozyilmaz, Jong-Hyun Ahn, Byung Hee Hong and Sumio Iijima, Nature Nanotechnology **5**, 574 (2010).

2. Walt A. de Heer, Claire Berger, Ming Ruan, Mike Sprinkle, Xuebin Li, Yike Hu, Baiqian Zhang, John Hankinson, and Edward Conrad, PNAS **41**, 16900 (2011).

3. S.S. Tinchev, Applied Surface Science **258**, 2931(2012).

4. www.SRIM.org

5. I.A. Faizrakhmanov, V.V. Bazarov, V.A. Zhikharev, A.L. Stepanov, I.B. Khaibullin, Vacuum **62,** 15 (2001).

6. A. C. Ferrari, and J. Robertson, Physical Review B **61**, 14 095 (2000).

7. T. Gokus, R.R. Nair, A. Bonetti, M. Boehmler, A. Lombardo, K.S. Novoselov, A.K. Geim, A.C. Ferrari, ACS Nano **3,** 3963 (2009).

8. C.N.R. Rao, K.S. Subrahmanyam, H.S.S. Ramakrishna Matte, B. Abdulhakeem, A. Govindaraj, B. Das, P. Kumar, A. Ghosh, D.J. Late, Sci. Technol. Adv. Mater. **11**, 054502 (2010).

9. Andrea C. Ferrari, Solid State Communications **143**, 47 (2007).